# Ensemble-CVDNet: A Deep Learning based End-to-End Classification Framework for COVID-19 Detection using Ensembles of Networks

Coşku Öksüz[a,*], Oğuzhan Urhan[b], Mehmet Kemal Güllü[c]

[a]Department of Electronics and Automation, Bozkurt Vocational School of University of Kastamonu, Kastamonu 37680, Turkey. coksuz@kastamonu.edu.tr

[b]Department of Electronics and Telecommunication Engineering, University of Kocaeli, Kocaeli 41380, Turkey. urhano@kocaeli.edu.tr

[c]Department of Electrical and Electronics Engineering, University of Bakırçay, Seyrek Campus, Menemen, Izmir 35665, Turkey. kemal.gullu@bakircay.edu.tr

[*]Corresponding author: coksuz@kastamonu.edu.tr; coskuoksuz@gmail.com

**Abstract**

The new type of coronavirus disease (COVID-19), which started in Wuhan, China in December 2019, continues to spread rapidly affecting the whole world. It is essential to have a highly sensitive diagnostic screening tool to detect the disease as early as possible. Currently, chest CT imaging is preferred as the primary screening tool for evaluating the COVID-19 pneumonia by radiological imaging. However, CT imaging requires larger radiation doses, longer exposure time, higher cost, and may suffer from patient movements. X-Ray imaging is a fast, cheap, more patient-friendly and available in almost every healthcare facility. Therefore, we have focused on X-Ray images and developed an end-to-end deep learning model, i.e. Ensemble-CVDNet, to distinguish COVID-19 pneumonia from non-COVID pneumonia and healthy cases in this work. The proposed model is based on a combination of three lightweight pre-trained models SqueezeNet, ShuffleNet, and EfficientNet-B0 at different depths, and combines feature maps in different abstraction levels. In the proposed end-to-end model, networks are used as feature extractors in parallel after fine-tuning, and some additional layers are used at the top of them. The proposed model is evaluated in the COVID-19 Radiography Database, a public data set consisting of 219 COVID-19, 1341 Healthy, and 1345 Viral Pneumonia chest X-Ray images. Experimental results show that our lightweight Ensemble-CVDNet model provides 98.30% accuracy, 97.78% sensitivity, and 97.61% $F_1$ score using only 5.62M parameters. Moreover, it takes about 10ms to process and predict an X-Ray image using the proposed method using a mid-level GPU. We believe that the method proposed in this study can be a helpful diagnostic screening tool for radiologists in the early diagnosis of the disease.

## 1. Introduction

The new type of coronavirus disease that started in Wuhan, China in December 2019 continues by affecting the whole world. The virus was first named 2019-novel coronavirus (2019-nCoV). On February 11, 2020, the disease was named COVID-19 by the World Health Organization (WHO). The virus that causes the disease is named *Severe Acute Respiratory Syndrome Coronavirus-2* (SARS-CoV-2) by ICTV (International Committee on Taxonomy of Viruses) [1]. WHO declared the new type of coronavirus disease as a global pandemic on March 11, 2020 [2]. The new type of coronavirus has a high degree of contagiousness and threatens the whole world with increasing cases and deaths. According to given records by WHO, as of December 5, 2020, there are 65,007,974 confirmed cases worldwide, while the number of deaths is 1,507,018. The dashboard reported by WHO given in Fig. 1 shows the weekly change of confirmed cases and draws attention to the rapid spread of the disease [3].

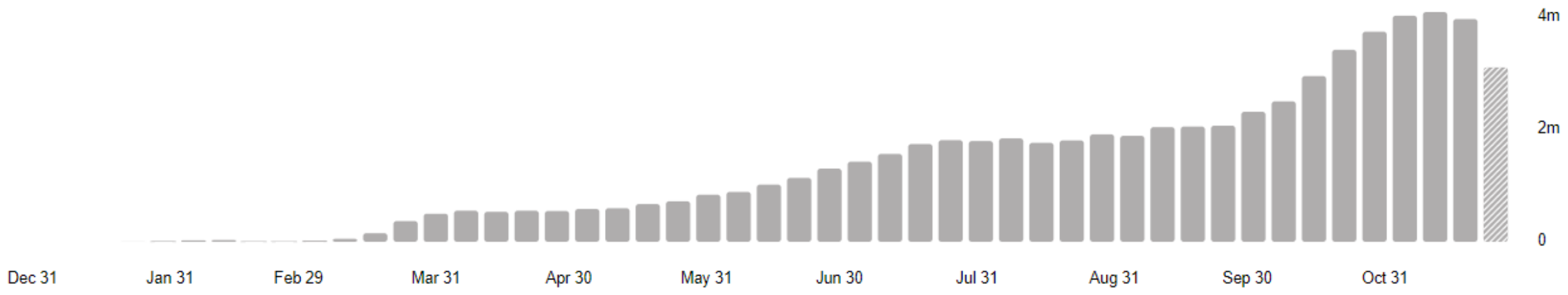

**Fig. 1.** The weekly change of confirmed worldwide cases for COVID-19 disease [3].

As of December 2, 2020, the draft landscape of COVID-19 candidate vaccines was shared by WHO [4]. According to the published document, there are 51 candidate vaccines in clinical evaluation and 163 candidate vaccines in pre-clinical evaluation. On the other hand, as of December 3, 2020, a COVID-19 vaccine has been approved by the UK [5]. However, even when the vaccines are approved by the regulators, it may take time to reach most of the population. Another concern is the thought that the virus will mutate, as the current vaccines under development are based on the known version of the virus. In September 2020, a new variant of COVID-19 was found in 12 people in Northern Jutland, Denmark. This variant was also found in a mink-rearing facility. For this reason, 17 million minks were culled on the thought that these mutations would adversely affect the deployment of vaccines [6].

In [7], the transmission modes of SARS-CoV-2 are indicated as contact, droplet, airborne, fomite, fecal-oral, bloodborne, mother-to-child, and animal-to-human. However, it is stated that the transmission of SARS-CoV-2 is mainly via droplets, and close contact with infected symptomatic cases [7], [8]. On the WHO site [9], the most common symptoms of COVID-19 are noted as fever, dry cough, and fatigue, while the most serious symptoms are difficulty breathing or shortness of breath, chest pain, and loss of speech and movement. In [10], a study was conducted that demonstrate the importance of initial symptoms for the elimination of COVID-19 disease or selecting patients for further diagnostic testing. Results from this study show that at least half of the participants with COVID-19 disease have a cough, sore throat, high fever, muscle or joint pain, fatigue, or headache. However, it is concluded that cough and sore throat are also common in people without COVID-19, and as a result, these symptoms alone are less helpful for diagnosing COVID-19. On the other hand, it is concluded that high fever, muscle pain, fatigue, and headache significantly increase the likelihood of COVID-19 disease when they are present.

In the current clinical routine, RT-PCR testing is regarded as the gold standard screening tool for identifying COVID-19 patients by WHO [8]. However, RT-PCR tests may miss positive cases due to problems such as laboratory-related errors, insufficiency of the viral material in the sample, improper removal of nucleic acid from the material, contamination and other technical problems. All these reduce the sensitivity of the RT-PCR test [1], [11], [12]. In [13], RT-PCR test of upper respiratory tract samples taken from hospitalized COVID-19 patients is performed to determine sensitivity. It is determined that the sensitivity of a single RT-PCR test is 82.2%, and when the test is repeated it increased to 90.6%. The other important thing worth mentioning is that the RT-PCR test is invasive and requires the intervention of a physician to take the nasal, and throat swabs. Due to the lack of specialist equipment and specialist health staffs, specimens taken from patients, especially in underpopulated places, are sent to laboratories in city centers. Hence, as stated in [8], since RT-PCR testing is time-consuming, requires expensive equipment, and requires biosafety conditions, it is not suitable for point-of-care diagnosis. On the other hand, it is revealed that radiological images acquired by techniques such as Chest Radiograph (X-Ray) and Computed Tomography imaging (CT) have important clues in the detection of the disease. Thus, radiologists have a critical role in detecting the disease at an early stage utilizing X-Ray and CT images [14]. The early detection of disease is indispensable to prevent the spread of SARS-COV-2. However, it is stated that the sensitivity to detect COVID-19 from X-ray images in the early stages of the disease is low [1]. As the disease progresses, multiple non-transparent (opaque) areas become evident on the X-Ray. These opacities turn into a whitened-lung image in the later stages of the disease. In [15], the clinical utility of X-Ray is examined, and it is concluded that X-Ray performed between the $6^{th}$ and $10^{th}$ days from the symptom onset is a more important predictor of severe disease than the X-Ray performed earlier. On the contrary, it is stated that detection sensitivity is high in the early stage of the disease with CT imaging [1].

In recent years, it is aimed to provide computer-aided diagnosis (CAD) systems that can support experts in their decisions in the medical field. The existence of these systems can also be a beneficial tool in underpopulated places where a specialist physician is not available. Such systems are designed based on machine learning methods. Deep learning which is a sub-field of machine learning differs from the classical machine learning approaches in which the important features (radiomics) for the task are manually explored, extracted and classified. In deep learning, the features to be extracted from the data are determined automatically. Thus, end-to-end systems in which feature extraction and classification are integrated are obtained. It is possible to categorize the proposed CAD systems for COVID-19 detection in the literature into two categories as X-Ray-based, and CT-based methods. Although it is hard to see the disease with a chest X-ray by a specialist in the early stages, deep learning methods are capable of learning hidden patterns (deep learning-based radiomics) in the data. In [16], a deep learning-based method using X-Ray, and CT images is proposed for the early detection of COVID-19.

Since the time the disease started and turned into a pandemic, much effort has been made by researchers to detect COVID-19 on radiological images. In [17], authors inspired by the DarkNet-19 model, the backbone of the YOLO (You Only Look Once) object detector model. The designed X-Ray based model is named as DarkCovidNet. An accuracy of 87.02% and 98.08% is achieved by DarkCovidNet for multi-class and binary classification tasks, respectively. In [18], a wide range of methods are used to detect COVID-19 disease. The SVM classification with extracted features by pre-trained ImageNet [19] models, fine-tuning of pre-trained models, designing a new deep learning model are the techniques used in this study. The ResNet50 features and SVM based classification yields the best performance with 94.7% accuracy. In [20], the pre-trained models, ResNet18, ResNet50, SqueezeNet, and DenseNet121 are used with transfer learning approach. It is stated that the sensitivity and specificity scores obtained with each model are around 98% and 90%, respectively. In [21], the authors designed a 5-layer convolutional network instead of transfer learning. After the training, the trained network is used as a feature extractor, and classical machine learning methods such as k-NN, SVM, and decision trees are used. In this study, 98.98% of accuracy is obtained with the SVM model. In [22], a custom CNN architecture with 15M parameters is proposed to learn COVID-19 related latent features. The overall accuracy is 97.94% in this work. In [23], the pre-trained ResNet-50 model is fine-tuned in three stages, using images of 128×128×3, 224×224×3, and 299×299×3 at each stage, respectively. The proposed method is named as COVID-ResNet. The accuracy obtained with the COVID-ResNet method with 25.6M parameters is 96.23%. In [24], a custom deep learning model with 11.75M parameters, COVID-Net is proposed for COVID-19 detection. The accuracy of 93.3% is achieved with the proposed model. In [25], a custom model for COVID-19 detection, COVID-AI detector is proposed. The obtained accuracy with COVID-AI is 90.5%. In [26], another deep learning-based approach, namely DeTraC (Decompose, Transfer, and Compose) is proposed. When using

DeTraC with the pre-trained ResNet18 model, the best accuracy score is achieved as 95.12%. In [27], a deep learning-based method consisting of three-stages is proposed to detect COVID-19 disease. In the first stage, it is determined whether an X-ray belongs to a normal or pulmonary case, while in the second stage, it is aimed to distinguish COVID-19 and generic pulmonary cases. In the third stage, it is aimed to visualize suspected areas in the X-Ray. An accuracy of 97% is achieved using the fine-tuned VGG-16 model in this work. In [28], a residual network architecture is proposed to detect local and global features for COVID-19 detection where 96.69% accuracy is obtained in this study.

The correct diagnosis ensures that people are not unnecessarily tested, unnecessarily loaded with medication, unnecessarily isolated, and get the right treatment quickly, reducing the risk of the spread of COVID-19. This is extremely important for individuals and saves time and resources. In this study, it is aimed to propose a highly sensitive deep learning-based rapid screening method that can be applied in all healthcare facilities to distinguish COVID-19 from healthy and viral pneumonia cases. For this purpose, it is aimed to develop the method based on X-Ray imaging, because it is an easy, fast, cheap, widely used technique, and therefore available in almost all healthcare facilities. Besides, it can also be said that X-Ray imaging is more patient-friendly than CT imaging due to the lower effective radiation doses applied to a patient as given in [29]. In order to obtain a model with low complexity, only the lightweight pre-trained models at different depths are used in combination in this work. SqueezeNet [30], ShuffleNet [31], MobileNet-v2 [32], and EfficientNet-B0 [33] are preferred in this work as the light-weight models where each of these models are configured to extract features in different abstraction levels. Hence, generated feature maps by a deep model are more abstract maps than a shallow model. Although this abstract information is distinguishing in classification performance, the lack of semantic information may reduce performance. Therefore, a classification scheme that combines model feature maps is adopted where the SqueezeNet is used to generate fine-grained maps, ShuffleNet is used to generate coarse-grained maps, and EfficientNet-B0 is used to generate coarser-grained maps. Thus, it is possible to combine fine-grained and coarse-grained features and to compensate for the loss of information in the deep model used alone. In summary, the main contributions of this paper are the followings:

- An X-Ray based lightweight end-to-end deep learning approach is proposed for COVID-19 detection.
- In order to compensate for the loss of information, three lightweight pre-trained models at different depths are used as feature extractors after fine-tuning.
- It only takes 10ms to process an X-Ray image with the designed end-to-end deep model. Therefore, it can be used as a rapid screening tool.

The study is organized as follows. In Section 2, the proposed classification framework that includes feature extraction, feature map fusion, and classification is introduced. Experimental results are presented in Section 3. Section 4 is devoted to discussions. Finally, Section 5 is devoted to conclusions.

## 2. Proposed Method

In this work, a deep learning-based classification scheme that combines networks with different depths is proposed to classify COVID-19, Normal, and Viral Pneumonia cases. The proposed classification framework is depicted in detail in Fig. 2. The proposed framework consists of feature extraction layers and some additional layers at the top of them. As seen in Fig. 2, three lightweight ImageNet [19] models, i.e. SqueezeNet [30], ShuffleNet-v2 [31], and EfficientNet-B0 [33], are used as feature extractors in parallel after fine-tuning. Networks at different depths generate deep activation maps at different abstraction-levels. For this purpose, the fine-tuned SqueezeNet with 22-layers-deep is used for fine-grained-abstract feature map generation. Other fine-tuned models, i.e. ShuffleNet with 50-layers-deep, and the EfficientNet-B0 with 82-layers deep, are used to generate coarse-grained, and coarser-grained-abstract feature maps, respectively. The activation maps obtained from each model are fed into the depth concatenation layer, and then combined to compensate the loss of information originating from spatial size reduction along the previous layers due to the pooling operations. After the channel concatenation, the combined channels are fed into the additional layers to learn important features and to classify COVID-19, Normal, and Viral Pneumonia cases. The detailed description of the proposed method is given in the following sections.

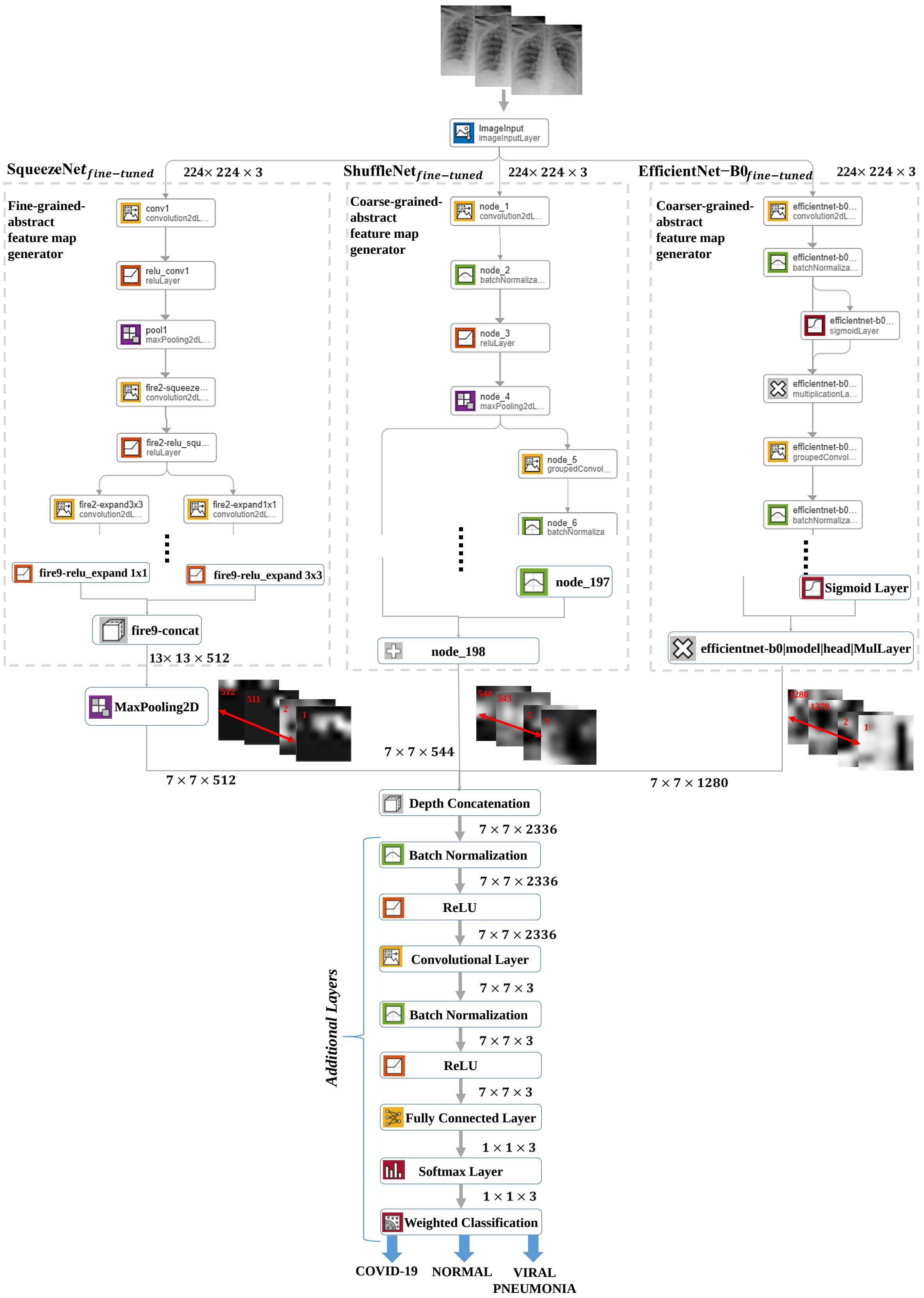

**Fig 2.** The proposed classification scheme.

### 2.1. Data Set

In the study, COVID-19 Radiography Database [12], which is a public data set in Kaggle, is used. This data set is well-curated from different resources, and consists of 2905 X-Ray images in Portable Network Graphics (PNG) file format. Each image is given as 1024×1024×3 in spatial resolution. There are three classes as COVID-19 (219 images), Normal (1341 images) and Viral Pneumonia (1345 images). The exemplary X-Ray images given in Fig. 3 belong to COVID-19, Normal, and Viral Pneumonia patients, respectively.

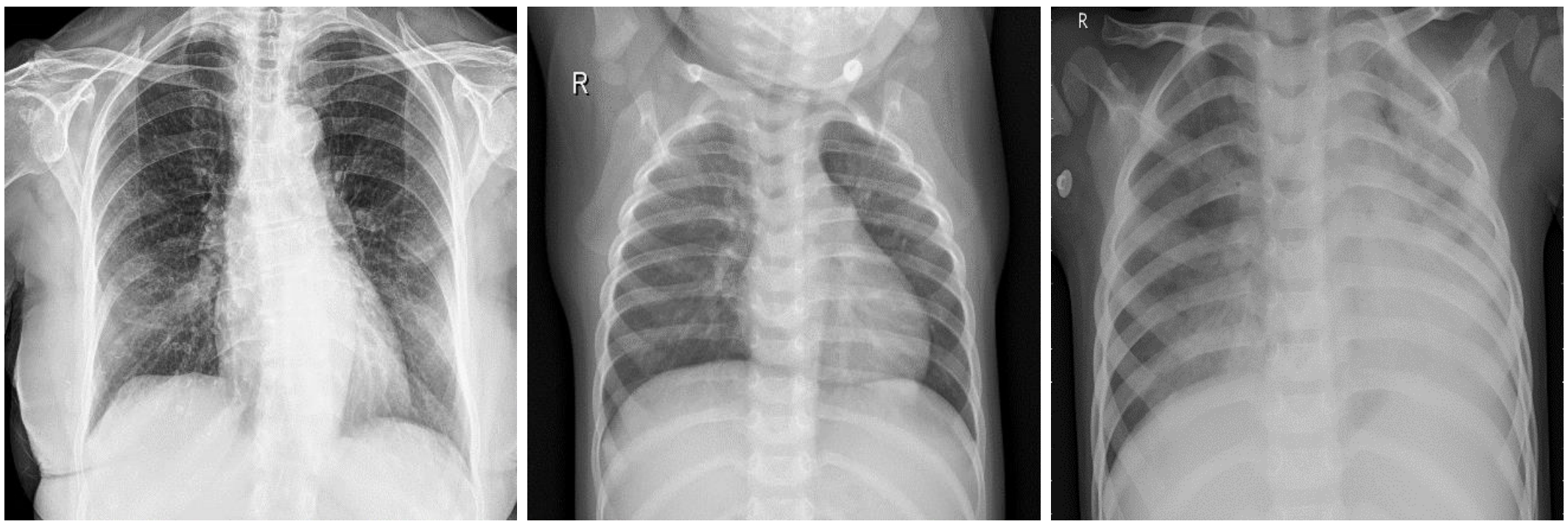

**Fig. 3.** Images from left to right are of COVID-19, Normal and Viral Pneumonia patients.

### 2.2. Fine-tuning and Feature Map Generation

Unlike classical machine learning, features are extracted without any prior knowledge in deep learning. The designed deep learning models in an end-to-end manner are trained to maximize a given objective function. This allows models to learn useful features that maximize classification performance for the task at the hand. However, one difficulty to learn such useful features in deep learning is the lack of medical data to train a CNN effectively. Fortunately, features learned by models that are trained with millions of data can be transferred to a new classification task thanks to the transfer learning [34]. One way of transfer learning is to use a pre-trained network directly as a feature extractor without fine-tuning. The other way is to fine-tune a pre-trained network. In [35], it is demonstrated that classification with features extracted after fine-tuning yields better generalization ability because of the learning of task-related features. Another point worth mentioning is that features are extracted in different abstract levels as the network deepens. Although the deep features are extremely important in classification, some useful information may be lost depending on the network depth. Therefore, in this work, in order to compensate for the information loss, we make use of three lightweight pre-trained models with different depths, i.e. SqueezeNet [30], ShuffleNet [31], and EfficientNet-B0 [33]. By combining these networks in parallel, an integrated feature map generator network is obtained.

First, the pre-trained models, SqueezeNet, ShuffleNet, and EfficientNet-B0, are fine-tuned for 10 epochs to support the weights for new task before building the proposed model. The subsections presented below explain how the feature map generators are derived using these fine-tuned networks.

#### *2.2.1. SqueezeNet as the feature map generator*

SqueezeNet [30] is a network with 18-layer depth and 1.24M parameters. The input spatial resolution of this network is 227×227×3. The main motivation behind the SqueezeNet approach was to design a network that yields high accuracy with a smaller number of parameters. Eventually, AlexNet [36] model accuracy is achieved on ImageNet [19] by SqueezeNet with 50× fewer parameters. It is worth noting that the input spatial resolution of this network is adjusted to 224 × 224 × 3 before fine-tuning for compatibility with other networks. For the fine-tuned model SqueezeNet, all the layers that come after the depth concatenation layer, i.e. fire9-concat, are removed. The activation maps obtained from fire9-concat layer are 13×13×512 in spatial resolution in this stage. A max-pooling layer with stride [2 2] and pooling size [2 2] is connected at the top of this layer. Finally, a feature map generator block is obtained that gives 512 activation maps in 7×7 spatial resolutions. We utilized this remaining architecture as our fine-grained-abstract feature map generator.

### *2.2.2. ShuffleNet as the feature map generator*

ShuffleNet [31] is a lightweight CNN architecture that utilizes point-wise grouped convolution and channel shuffle operation. ShuffleNet is specifically designed for mobile devices. This network has 50-layer depth and 1.4M parameters. The input spatial resolution of the network is 224×224×3. For the fine-tuned model ShuffleNet, all the layers that come after the element-wise addition layer, node_198, are removed. The remaining architecture is used as our coarse-grained-abstract feature map generator. A total of 544 activation maps in spatial resolution 7×7 are obtained from this feature map generator block.

### *2.2.3. EfficientNet-B0 as the feature map generator*

In EfficientNet [33], it is revealed that carefully balancing network depth, width, and resolution improves classification performance. Efficient networks presented in [33] consist of eight successive models between B0-B7 by considering these criteria. The EfficientNet-B0 is the base model with 82-layer depth and 5.3M parameters. The input spatial resolution of this network is 224×224×3. Other models are derived from the base model successively. The model complexity increases with each derivation. The input spatial resolution of each separate model is also increased to capture more fine-grained features. The highest ImageNet accuracy is achieved with the final model B7. In the study, the EfficientNet-B0 model is used because of its low complexity. For the fine-tuned model EfficientNet-B0, all the layers that come after the layer, efficientnet-b0|model|head|MulLayer, are removed from the network. Thus, the remaining architecture is used as our coarser-grained-abstract feature map generator. The number of 1280 activation maps in 7×7 resolution are obtained from this feature map generator block.

## 2.3. Building the proposed network

This section introduces the building process of the proposed network given in Figure 2. The inputs of the feature map generator blocks are linked together to have a single-input network. In this way, three separate feature map generator blocks (surrounded by the gray dashed lines in Fig. 2) that take input in 224 × 224 × 3 spatial resolutions and process in parallel are obtained.

### *2.3.1. Feature map fusion*

The feature maps generated by fine-grained, coarse-grained, and coarser-grained feature map generators are fed to the depth concatenation layer. This operation enables to combine feature maps that are in different abstract levels. Thus, in a manner, feature maps are aligned from fine to coarse. Finally, 2336 activation maps in 7×7 spatial resolutions are obtained from the output of the depth concatenation layer. At the end of this step, the feature map generator backbone of the proposed network is completed. The weights of the backbone are frozen to avoid recalculations of gradients while the proposed model is being trained.

### *2.3.2. Additional Layers*

The additional layers consist of Batch Normalization, ReLU, Convolution, Batch Normalization, ReLU, Fully Connected, Softmax, and Weighted Classification layers that come after another. First, the tensor in size 7×7×2336 obtained after the feature map fusion is travelled into Convolutional Layer by following Batch Normalization, and ReLU layers. The three kernels with stride $[1 \quad 1]$ are used in the Convolution Layer to reduce the number of parameters. As a result, the obtained tensor at the output of this layer is 7×7×3 in size. Then, the tensor is travelled into the Fully Connected Layer by following Batch Normalization, and ReLU layers, respectively. Since the number of activation elements are equal to the number of classes, the resulting final CNN codes are 1×1×3 in size. The CNN codes are fed to the Softmax Layer to obtain class prediction scores. Finally, the weighted classification layer is used after the Softmax layer because there is highly class imbalance in data set. The weighted cross entropy loss is computed at this layer for classification task. At the end of all these steps, a single-input and single-output deep learning model is built.

## 3. Experiments and Results

### 3.1. Implementation

The experiments are performed on a computer with 2.6 GHz CPU, 16 GB RAM, and NVIDIA GeForce RTX 2070 GPU and all the experiments were carried out using the MATLAB® R2020b software package.

### 3.2. Hyperparameters and Data Augmentation

As stated in Section 2.2, the pre-trained models, SqueezeNet, ShuffleNet, and EfficientNet-B0 are fine-tuned to support the new task before building the proposed model. The training data is augmented at each epoch. The training hyperparameters are given in Table 1 for each model. The training time for models, SqueezeNet, ShuffleNet, and EfficientNet-B0 are about 4, 8 and 21 minutes, respectively. While building Ensemble-CVDNet, the fine-tuned EfficientNet-B0 (EfficientNet-B0$_{ft}$) model is first combined with the fine-tuned SqueezeNet (SqueezeNet$_{ft}$). The network formed as a result of this combination is named as Ensemble-CVDNet #1. The training time for Ensemble-CVDNet#1 is about 8 minutes. After that, the proposed model is built by integrating the fine-tuned ShuffleNet (ShuffleNet$_{ft}$) model to Ensemble-CVDNet #1. Training time for the proposed model is about 10 minutes. The proposed model is named as Ensemble-CVDNet #3 in the experiments.

The MobileNet-v2 [32] is a lightweight model with 53-layer depth and 3.5M of parameters. In the experiments, while constructing the proposed model, MobileNet-v2 is used instead of the ShuffleNet model as well, as it is comparable in terms of the model depth. This model is named as Ensemble-CVDNet #2. The training time for the fine-tuned MobileNet-v2 (MobileNetv2$_{ft}$) is about 11 minutes and the training time for Ensemble-CVDNet #2 with MobileNet-v2 is about 12 minutes.

**Table 1**
The training hyperparameters and data augmentation techniques are used in this study.

| SETTINGS | | | |
|---|---|---|---|
| **HYPERPARAMETER** | **SqueezeNet, ShuffleNet, MobileNet-v2, EfficientNet-B0** | **Ensemble-CVDNet #1, Ensemble-CVDNet #2, Ensemble-CVDNet #3** | **Data augmentation** |
| **Optimizer** | ADAM | ADAM | - Random reflection on both X and Y axis. |
| **Mini Batch Size (MBS)** | 32 | 32 | |
| **Max Epoch** | 10 | 10 | - Random rescaling between [0.75 1.25]. |
| **Global Learning Rate** | $10^{-4}$ | $10^{-3}$ | |
| **Validation Frequency** | 65 | 65 | - Random rotation between $[-30^{\circ} \quad 30^{\circ}]$. |
| **L2 Regularization** | $10^{-4}$ | $10^{-4}$ | |
| **Weight & Bias Learn Rate Factor** | 10 | - | - Random horizontal & vertical translation between [-3 3] pixels. |
| **Weight vector in classification layer** | [0.75 0.1 0.15] | [0.75 0.1 0.15] | |

## 3.3. Evaluation of the Proposed Classification Framework

The evaluation process of the proposed classification framework is depicted in Fig. 4. We utilized five-fold cross-validation. As seen in Fig 4, the entire data set is divided into five non-overlapping folds. While one fold is used as a test set, the remaining folds are used to train the model each time. During the training phase, 10% of the samples in the training set are randomly picked to control over-fitting. Once the model is trained, it is evaluated using the corresponding test fold. After this process is repeated five-times, confusion matrices obtained on each test fold are summed up to see the whole model performance.

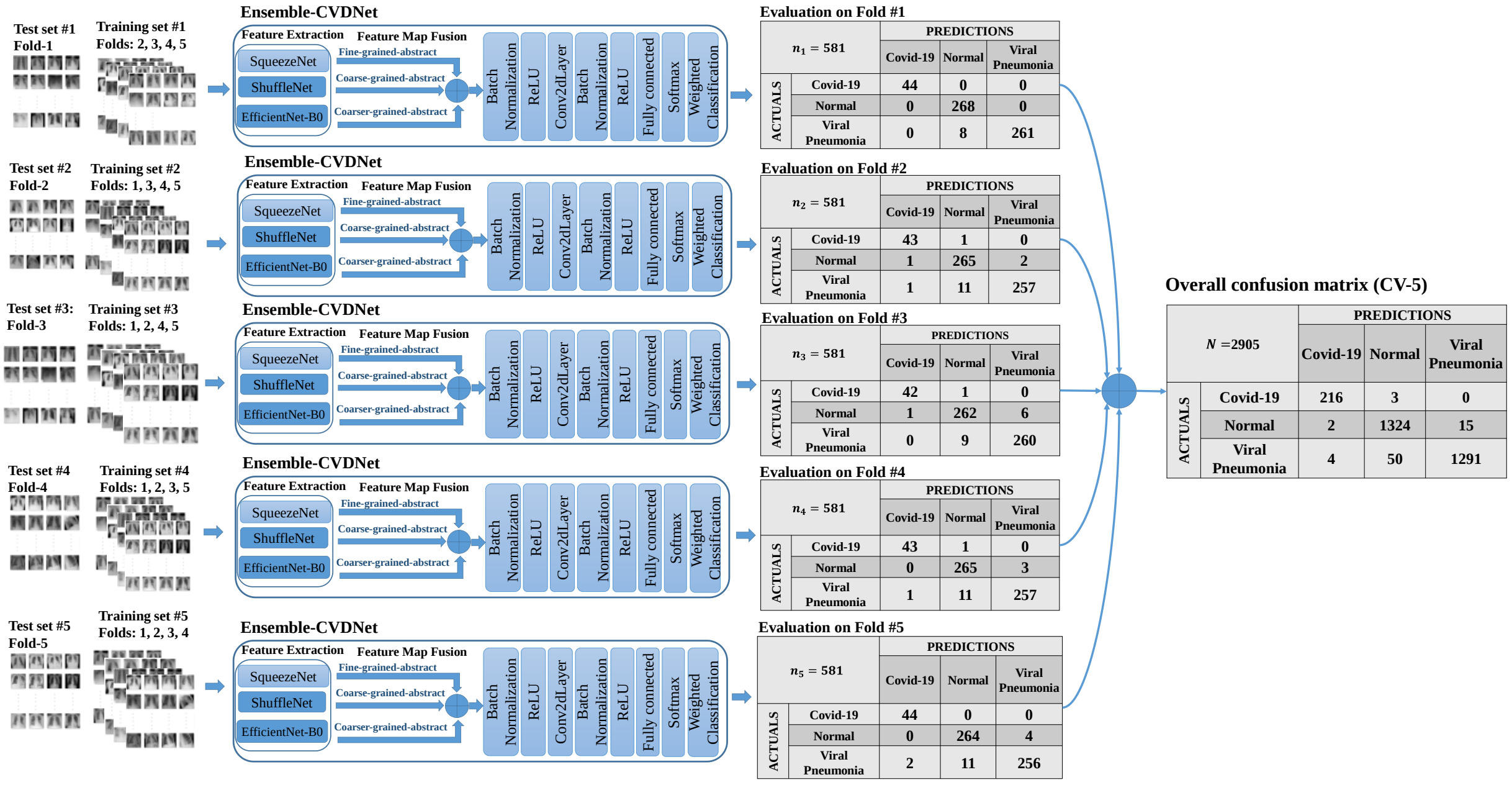

**Fig 4.** Evaluation of the proposed classification framework using five-fold cross-validation.

## 3.4. Performance metrics

In many studies, classification accuracy is used as a common metric to evaluate the performance of a classifier model. The classification accuracy (ACC) is computed using (1).

$$\mathrm{ACC}=\frac{\mathrm{TP}+\mathrm{TN}}{\mathrm{TP}+\mathrm{TN}+\mathrm{FP}+\mathrm{FN}} \tag{1}$$

However, using an imbalanced data set makes a classifier model biased towards the majority class. As a result, the model votes for the majority class most of the time against the minority class. Thus, using classification accuracy as an evaluation metric for a model trained using imbalanced data set will not reflect realistic results. The data set used in this study is imbalanced. The more realistic results can be obtained using metrics such as precision (PPV), recall (or sensitivity, TPR), specificity (SPEC), and $F_1$-score. Recall, precision, specificity and $F_1$ scores are computed using (2). The $F_1$ score is important since it represents both recall and precision scores as a single value.

$$\mathrm{TPR}=\frac{\mathrm{TP}}{\mathrm{TP}+\mathrm{FN}},\quad \mathrm{PPV}=\frac{\mathrm{TP}}{\mathrm{TP}+\mathrm{FP}},\quad \mathrm{SPEC}=\frac{\mathrm{TN}}{\mathrm{TN}+\mathrm{FP}},\quad F_1\text{-Score}=2\times\frac{\mathrm{TPR}\times\mathrm{PPV}}{\mathrm{TPR}+\mathrm{PPV}} \tag{2}$$

## 3.5. Performance Comparison

In this section, performance of the proposed model is compared with the other methods used in this work, and the literature. In Table 2, the obtained scores on each fold for each method is given with average cross-validation scores.

**Table 2**
The comparison of accuracy scores of each model obtained on each test fold. The subscript ft indicates that the model is fine-tuned

| Method | Accuracy (%) | | | | | |
|---|---|---|---|---|---|---|
| | **Fold-1** | **Fold-2** | **Fold-3** | **Fold-4** | **Fold-5** | ***Avg.*** |
| $\text{SqueezeNet}_{ft}$ | 94.84 | 94.26 | 95.64 | 95.52 | 96.21 | *95.30* |
| $\text{ShuffleNet}_{ft}$ | 97.25 | 94.72 | 95.64 | 97.25 | 97.13 | *96.40* |
| $\text{MobileNet-v2}_{ft}$ | 98.39 | 96.90 | 96.90 | 97.59 | 97.25 | *97.41* |
| $\text{EfficientNet-B0}_{ft}$ | 98.74 | 97.59 | 96.56 | 96.90 | 97.36 | *97.43* |
| $\text{SqueezeNet}_{ft}$ & $\text{EfficientNet-B0}_{ft}$ (Ensemble-CVDNet #1) | 98.62 | 97.36 | 97.02 | 97.48 | 97.82 | *97.66* |
| $\text{SqueezeNet}_{ft}$ & $\text{MobileNet-v2}_{ft}$ & $\text{EfficientNet-B0}_{ft}$ (Ensemble-CVDNet #2) | 98.39 | 98.16 | 97.93 | 97.93 | 97.25 | *97.93* |
| $\text{SqueezeNet}_{ft}$ & $\text{ShuffleNet}_{ft}$ & $\text{EfficientNet-B0}_{ft}$ (Ensemble-CVDNet #3) | **99.08** | **98.16** | **98.05** | **98.16** | **98.05** | ***98.30*** |

The best classification accuracy is obtained with the $\text{EfficientNet-B0}_{ft}$ model among the single architectures. On the other hand, $\text{SqueezeNet}_{ft}$ performs worst compared to $\text{ShuffleNet}_{ft}$, $\text{MobileNet-v2}_{ft}$, and $\text{EfficientNet-B0}_{ft}$. Additionally, $\text{ShuffleNet}_{ft}$ underperforms compared to $\text{MobileNet-v2}_{ft}$ and $\text{EfficientNet-B0}_{ft}$. All these indicate that more detailed features are extracted by deeper networks. In order to show the effect of the $\text{SqueezeNet}_{ft}$ on the performance of the proposed model, the created transition model i.e. Ensemble-CVDNet #1, leads to an increase of about 0.23% on the average cross-validation accuracy. After that, the model formed by integrating the $\text{MobileNet-v2}_{ft}$ to the Ensemble-CVDNet#1, i.e. the Ensemble-CVDNet #2, contributes about 0.27% to the avgerage cross-validation accuracy. When it is compared with the EfficientNet-B0, Ensemble-CVDNet #2 contributes about 0.5% to average cross-validation accuracy. Finally, the proposed model (Ensemble-CVDNet #3) is built by integrating the $\text{ShuffleNet}_{ft}$ to Ensemble-CVDNet #1. As seen in Table 2, this integration provides about 0.64% to the average cross-validation accuracy. In this way, the obtained cross-validation accuracy with the single $\text{EfficientNet-B0}_{ft}$ model is increased by about 0.87% by Ensemble-CVDNet #3. The best cross-validation accuracy is achieved with the proposed method. The confusion matrix obtained on each fold is given in Fig. 2 for the proposed method.

**Table 3**
TPR, PPV, SPEC, $F_1$-Score, and ACC scores of the methods used in this study. Results are given with 95% confidence interval.

| Method | Avg. TPR (%) | Avg. PPV (%) | Avg. SPEC (%) | Avg. $F_1$ Score (%) | Avg. ACC (%) |
|---|---|---|---|---|---|
| $\text{SqueezeNet}_{ft}$ | 89.56 ±1.11 | 93.88 ±0.87 | 95.69 ±0.73 | 91.67 ±1.0 | 95.30 ±0.76 |
| $\text{ShuffleNet}_{ft}$ | 94.71 ±0.81 | 95.71 ±0.73 | 96.70 ±0.64 | 95.21 ±0.77 | 96.40 ±0.67 |
| $\text{MobileNet-v2}_{ft}$ | 95.28 ±0.77 | 96.80 ±0.64 | 97.62 ±0.55 | 96.04 ±0.71 | 97.41 ±0.57 |
| $\text{EfficientNet-B0}_{ft}$ | 96.46 ±0.67 | 96.63 ±0.65 | 97.66 ±0.54 | 96.55 ±0.66 | 97.43 ±0.57 |
| Ensemble-CVDNet #1 | 96.96 ±0.62 | 96.65 ±0.65 | 97.89 ±0.52 | 96.81 ±0.63 | 97.66 ±0.54 |
| Ensemble-CVDNet #2 | 97.26 ±0.59 | 97.29 ±0.59 | 98.12 ±0.49 | 97.28 ±0.59 | 97.93 ±0.51 |
| Ensemble-CVDNet #3 **(Proposed method)** | **97.78** ±0.53 | **97.43** ±0.57 | **98.48** ±0.44 | **97.61** ±0.55 | **98.30** ±0.47 |

In Table 3, recall, specificity, precision, $F_1$-score, and accuracy scores are given for each method. Classifying a patient with COVID-19 as Normal or viral pneumonia (False Negatives) is a more serious mistake than classifying a normal or viral pneumonia patient as a COVID-19 (False Positives) patient. This may lead to the rapid spread of SARS-COV-2, especially when its contagiousness is considered. Therefore, a COVID-19 classifier needs to have a high hit rate (or recall score-TPR-). As seen in Table 3, better recall scores are achieved with Ensemble-CVDNet#1, Ensemble-CVDNet#2, and Ensemble-CVDNet#3 compared to the single models. Similarly, better specificity scores (SPEC) are achieved with Ensemble-CVDNet#1, Ensemble-CVDNet#2, and Ensemble-CVDNet#3 compared to the single models. On the other hand, the best scores for each metric are achieved by the proposed Ensemble-CVDNet#3 method. All these indicate that ensembling the networks compensates the information loss, and reduces false positives and negatives. This is also shown in Table 4where the confusion matrix given for EfficientNet-B0, Ensemble-CVDNet #1, and Ensemble-CVDNet #3 demonstrates how the errors are distributed among classes. As seen in Table 4, the total cost of misclassification is first reduced to 102 from 112 with Ensemble-CVDNet #1, and then it is reduced to 74 from 102 with Ensemble-CVDNet #3.

**Table 4**
Confusion matrix belongs to EfficientNet-B0, Ensemble-CVDNet #1, and Ensemble-CVDNet #3. The order is EfficientNet-B0/Ensemble-CVDNet #1/Ensemble-CVDNet #3.

| **N=2905** | | **PREDICTED CLASS** | | |
|---|---|---|---|---|
| | | **COVID-19** | **NORMAL** | **VIRAL PNEUMONIA** |
| **TRUE CLASS** | **COVID-19** | 213/215/**216** | 6/4/**3** | 0/0/**0** |
| | **NORMAL** | 3/2/**2** | 1314/1322/**1324** | 24/17/**15** |
| | **VIRAL PNEUMONIA** | 2/5/**4** | 77/74/**50** | 1266/1266/**1291** |

The scatter plot given in Fig. 5 shows the number of parameters versus the average accuracy scores of different methods compared. The fine-tuned single architectures are denoted by pentagram-shaped markers in the scatter plot, while the ensemble ones are denoted by hexagram-shaped markers. As seen in Fig 5., the $\text{ShuffleNet}_{ft}$ increases accuracy score by about 1.1% with an additional 0.14M of parameters compared to the $\text{SqueezeNet}_{ft}$. Similarly, $\text{MobileNet-v2}_{ft}$ with an additional 1.38M of parameters compared to the $\text{ShuffleNet}_{ft}$ contributes about 1% to the accuracy. However, an additional 1.79M of parameters for $\text{EfficientNet-B0}_{ft}$ compared to $\text{MobileNet-v2}_{ft}$ bring no significant gain (about 0.02%) in accuracy. On the other hand, accuracy scores obtained with fine-tuned single models can be increased by ensemble models, as indicated by hexagram-shaped markers in Fig. 5. Accordingly, the Ensemble-CVDNet #1 built by bringing an additional 0.72M of parameter load to $\text{EfficientNet-B0}_{ft}$, makes an increase in accuracy score about 0.23%. Moreover, Ensemble-CVDNet #3 built by incorporating $\text{ShuffleNet}_{ft}$ to Ensemble-CVDNet #1 increases the accuracy about 0.64% compared to Ensemble-CVDNet #1 by bringing an additional 0.86M of parameter load. When $\text{MobileNet-v2}_{ft}$ is used as a coarse-grained feature extractor of the designed classification scheme (i.e. Ensemble-CVDNet #2), the accuracy achieved stays behind the Ensemble-CVDNet#2 with an additional 1.38M of parameters compared to Ensemble-CVDNet #3. The main reason for this is that some feature maps extracted by $\text{MobileNet-v2}_{ft}$ and $\text{EfficientNet-B0}_{ft}$ are quite similar. This causes redundant features to come into play which degrades classification performance. Consequently, in this study, the best performance is obtained Ensemble-CVDNet #3 with 5.62M of parameters, which is also lightweight. The additional 1.58M of parameters (Total number of parameters of $\text{SqueezeNet}_{ft}$ and $\text{ShuffleNet}_{ft}$) bring 0.87%, 1.32%, 0.8%, 0.82%, and 1.06% of gain in ACC, TPR, PPV, SPEC, and $F_1$ scores compared to $\text{EfficientNet-B0}_{ft}$, respectively.

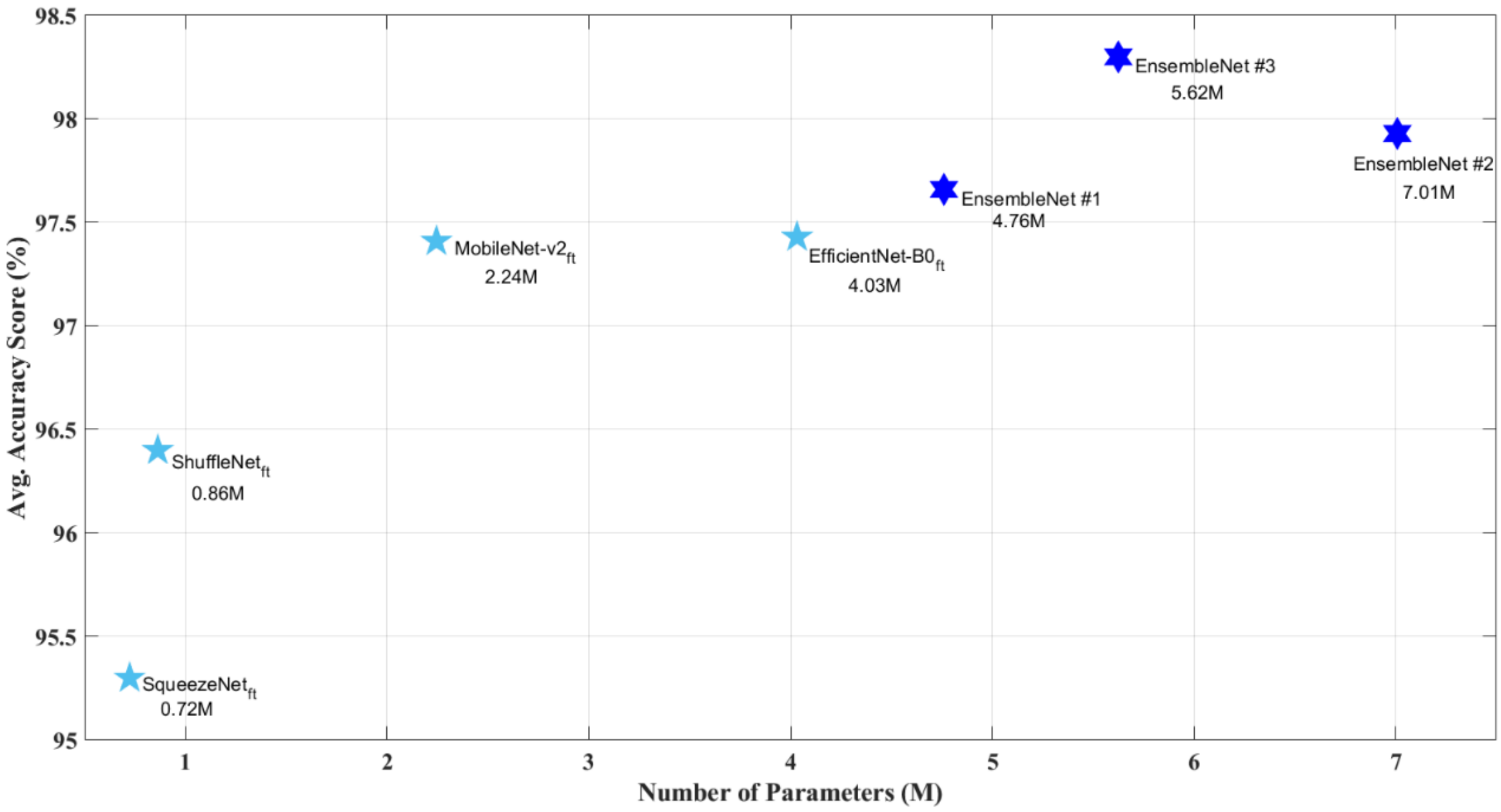

**Fig 5.** The scatter plot shows the number of parameters versus the accuracy scores achieved.

The average inference time for each method is also evaluated. For this purpose, the best performing models in the cross-validation are used for each method. X-Ray images in the corresponding test fold are fed 10-times to obtain consistent results and response times are recorded and averaged. Accordingly, it takes about 10.3ms to process and classify an input X-Ray image with the proposed method. It takes about 7.8ms, and 10.1ms to process an image for Ensemble-CVDNet #1, and Ensemble-CVDNet #2, respectively. For fine-tuned models, SqueezeNet$_{ft}$, ShuffleNet$_{ft}$, MobileNet-v2$_{ft}$, and EfficientNet-B0$_{ft}$, it takes about 7.5ms, 7.8ms, 7.7ms, and 7.9ms, respectively.

### 3.6. Visualization

In this section, it is aimed to reveal which parts of the image affect the classification decisions made by the proposed model in COVID-19 detection. It is also aimed to compare the decisions made by the single model, EfficientNet-B0. For this purpose, the gradient-weighted class activation mapping (Grad-CAM) technique [37] is used. In this technique, the gradient of the final classification score with respect to the final activation map is used to explain which parts of the image are important in classification. For this purpose, the last ReLU layer of the proposed method, i.e. relu_2_2, where the activation maps are obtained is used. For EfficientNet-B0 model, the last convolutional layer, efficientnet-b0|model|head|conv2d|Conv2D, is used.

In Fig. 6, images at the first column belong to COVID-19 patients, while the images in the second and third column show GradCAM images returned by the EfficientNet-B0 and the proposed method, respectively. The X-Ray images in the first three rows are correctly classified as COVID-19 patient by each method. The X-Ray image in the fourth row is correctly classified as COVID-19 by the proposed method, whereas it is misclassified as Normal (False negative) by EfficientNet-B0. On the other hand, the X-Ray image in the fifth row is misclassified by both methods. In GradCAM images, the important areas in the classification are shown in warm colors, while less important areas are shown in cool colors. It is hard to interpret all these GradCAM images of each method without a medical expert. However, it is observed that the areas returned with the proposed method are generally within the lungs and focus more on some specific areas. Whereas, some irrelevant areas outside the lungs (for example images in the first and second rows) are returned by the EfficientNet-B0 model.

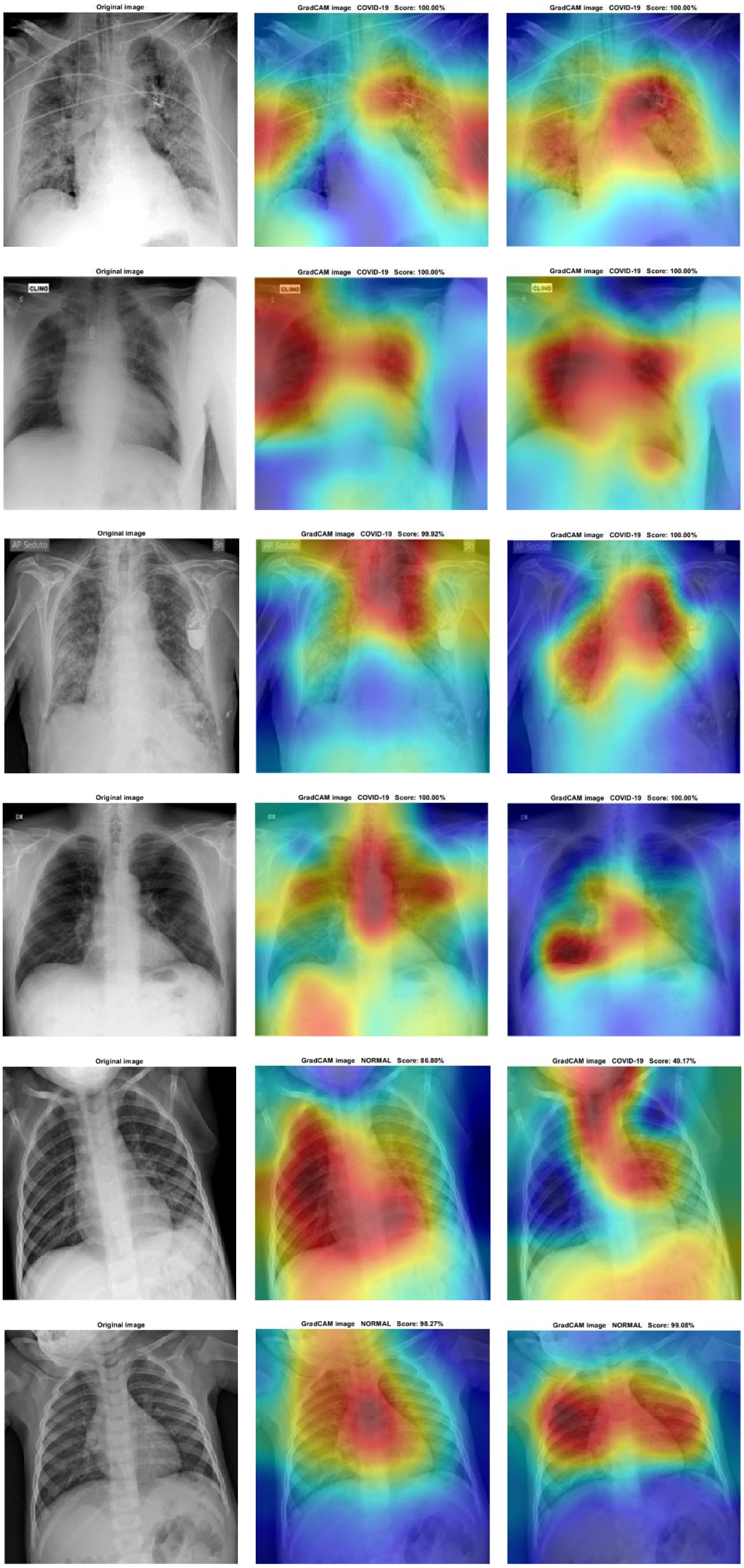

**Fig 6.** GradCAM visualizations. The first column shows original X-Ray images belong to COVID-19 patients. The other columns show GradCAM visualizations of the EfficientNet-B0, and the proposed method, respectively.

## 4. Discussion

Since the COVID-19 outbreak began and turned into a pandemic, intense efforts have been made by researchers to detect the disease from medical images. Many researchers have developed their methods using the public data set shared on the GitHub website by Joseph Paul Cohen [38]. Some other researchers have developed their methods using other private data sets, while others have developed methods using the data set collected from various sources. In this study, we developed our method using the public data set, the COVID-19 Radiography Database [12]. This data set consists of cases collected from Cohen's dataset, sample cases collected from the sirm.org website, cases from 43 publications, and data collected from the UCSD-Guangzhou pediatric data set [39] [40]. In Table 5 and Table 6, we compared our method with other X-Ray-based methods based on cross validation scores. We also give the results for the best fine-tuned model on the corresponding test set (fold). In Table 5, the performance of the proposed method is compared with the other X-Ray based methods using different data sets.

**Table 5**
Comparison of classification results with related studies.

| Works | Total parameter ≈ | $\frac{Covid19_{test}}{Total_{test}}$ | Avg. TPR (%) | Avg. PPV (%) | Avg. SPEC (%) | Avg. $F_1$ Score (%) | Avg. ACC (%) | Running time for one image ≈ |
|---|---|---|---|---|---|---|---|---|
| Ozturk et al. **[17]** | 1.164M | 107/624 | 95.13 | 98.03 | 95.3 | 96.51 | 98.08 | - |
| Ismael & Şengür **[18]** | 23.54M | 43/95 | 91.00 | - | 98.89 | 94.79 | 94.74 | 0.51s |
| Minaee et al. **[20]** | 0.72M | 98/3100 | 98.00 | - | 92.90 | - | - | - |
| Wang & Wong **[24]** | 11.75M | 100/300 | 93.33 | 93.56 | - | - | 93.30 | - |
| Mangal et al. **[25]** | 26M | 30/654 | 91.28 | 94.15 | 92.34 | 92.69 | 93.68 | - |
| Farooq & Hafeez **[23]** | 25.6M | 8/637 | 96.92 | 96.86 | - | 96.88 | 96.23 | - |
| Abbas et al. **[26]** | - | 285/529 | 97.91 | - | 91.87 | - | 95.12 | - |
| Karthik et al. **[22]** | 15.6M | 112/3054 | 97.54 | 96.34 | - | 96.90 | 97.94 | - |
| Brunese et al. **[27]** | 14.74M | 50/3326 | 91.50 | - | 96.00 | 91.5 | 97.00 | 5.07s |
| **Proposed method (CV-5)** | 5.62M | 219/2905 | 97.78 ±0.53 | 97.43 ±0.57 | 98.48 ±0.44 | 97.61 ±0.55 | 98.30 ±0.47 | 10.3ms |
| **Proposed method (The best fine-tuned model)** | | 44/581 | **99.01** ±0.8 | **99.03** ±0.8 | **99.15** ±0.75 | **99.02** ±0.8 | **99.08** ±0.77 | |

The running time is an important factor for COVID-19 detection methods when it is considered the density in healthcare facilities during the pandemic. The running time for the methods in [18], and [27] are reported, whilst it is not provided in other works. In [18], several methods have been used, but the best results are obtained using the fine-tuned ResNet-50 model features + SVM-based classification. The running time for this method is reported as 48.9s. Since the number of 95 images are used in the testing phase, we computed the running time of the method for one image as 0.51 seconds. Besides, very few samples are used in the testing phase. Only 8 samples from the COVID-19 class were used in [23], and the accuracy of COVID-19 detection is reported to be 100% that is why the average TPR score increases. All these situations hinder assessing the actual performance of the models. In [27], a classification scheme consisting of two separate deep models using the VGG-16 model is proposed. The first model is used to predict healthy and pneumonia cases, while the other model is used to predict disease and COVID-19 pneumonia cases. The VGG-16 network weights are quite large. As a result, the average prediction time for stage-1 and stage-2 models are 2.57s, and 2.50s, respectively. Accordingly, it takes about 5.06s to predict a COVID-19 patient. Some other heavy models are used in the works [22], [24], [25], [26]. Relatively good scores are reported with lightweight architectures presented in [17] and [20].

In Table 6, performance of the proposed method is compared with the methods using the COVID-19 Radiography Database. Considering the studies using the same data set as our study, although the accuracy reported with simple architecture in [21] is 98.97%, the average TPR score is relatively low. Relatively good scores were reported with the lightweight architecture in [28]. In [42], a hybrid classification framework consisting of 2D-Curvelet transform, CSSA (Chaotic Salp Swarm Algorithm), and EfficientNet-B0 is proposed. Accordingly, the designed classification scheme indicates that the X-Ray image of a suspected case is first transferred to the curvelet domain and then the resulting transformation domain feature map is fed to CNN for prediction. Although the state-of-the-art performance scores are reported, the learning of the transformation domain features by CNN is one of the drawbacks because of the transformation cost. There is no information presented about the running time for the proposed method in this work.

As seen in Table 6, the best results in this study are obtained with the proposed method. Considering both the CV-5 performance and the performance of the proposed method on the test set, it is seen that it gives consistent results based on all metrics. Moreover, it takes about 10.3ms to predict an X-Ray image with the model we proposed.

**Table 6**
Comparing the classification results with reference works using the COVID-19 Radiography Database.

| **Works** | **Total Parameters ≈** | **TPR (%)** | | | | **PPV (%)** | | | | **SPEC (%)** | | | | **$F_1$ (%)** | **ACC (%)** |
|---|---|---|---|---|---|---|---|---|---|---|---|---|---|---|---|
| | | **COVID-19** | **NORMAL** | **VIRAL PNEUMONI** | ***Average*** | **COVID-19** | **NORMAL** | **VIRAL PNEUMONI** | ***Average*** | **COVID-19** | **NORMAL** | **VIRAL PNEUMONI** | ***Average*** | ***Average*** | ***Average*** |
| Nour et al. **[21]** | 6.03M | - | - | - | 89.39 | - | - | - | - | - | - | - | 99.75 | 96.72 | 98.97 |
| Ouchicha et al. **[28]** | 5.31M | 97.2 | 96.7 | 96.6 | 96.84 | 95.9 | 96.6 | 96.9 | 96.72 | 99.7 | 97.1 | 97.4 | 98.03 | 96.68 | 96.69 |
| Altan & Karasu **[42]** | - | **100** | 99.4 | **99.6** | **99.69** | **100** | **99.6** | 99.4 | **99.69** | **100** | 99.8 | 99.7 | **99.84** | **99.69** | **99.79** |
| **Proposed** CV-5 | 5.62M | 98.6 ±0.42 | 98.7 ±0.41 | 96.0 ±0.71 | 97.78 ±0.53 | 97.3 ±0.59 | 96.2 ±0.69 | 98.9 ±0.38 | 97.43 ±0.57 | 99.8 ±0.16 | 96.6 ±0.66 | 99.0 ±0.36 | 98.48 ±0.44 | 97.61 ±0.55 | 98.30 ±0.47 |
| **Proposed** The best model | 5.62M | **100** ±0.0 | **100** ±0.0 | 97.0 ±1.38 | 99.01 ±0.8 | **100** ±0.0 | 97.1 ±1.36 | **100** ±0.0 | 99.03 ±0.79 | **100** ±0.0 | 97.4 ±1.29 | **100** ±0.0 | 99.15 ±0.74 | 99.02 ±0.8 | 99.08 ±0.77 |

The pros of the proposed method are as follows:

- It yields 100% sensitivity in the detection of COVID-19 cases.
- Tedious manual feature seeking process is bypassed using the end-to-end deep learning model.
- Thanks to the integration of three different models at different depths, it captures important patterns related to COVID-19 cases.
- It is a lightweight model with few parameters and offers a fast screening tool with the ability to diagnose COVID-19 in milliseconds.

The cons of the proposed method are as follows:

- Model is trained with limited number of COVID-19 cases, because there are still not enough COVID-19 cases in data sets. As the datasets are updated with increasing number of cases, more important features of COVID-19 can be learned by the model.
- The input spatial resolution of the method is set to 224 × 224 × 3 to reduce the number of parameters. This may cause some fine-grained features to be missed.

In future studies, we plan to re-validate our model when the number of cases in public data sets are updated. In addition, we plan to place our model in the cloud to assist doctors to promptly diagnose, isolate and treat COVID-19 patients. Thus, it will also be beneficial for health centers located in areas with low population, where there are not enough medical equipment and health personnel.

## 5. Conclusions

In this work, a lightweight CNN model that uses an ensemble of three lightweight models is proposed for COVID-19 detection using X-Ray images. Many studies use a wide variety of pre-trained ImageNet models with a fine-tuning strategy to determine which model is performing well for COVID-19 detection. In many other studies, designed networks are trained from scratch. Because of learning complex feature patterns, pre-trained models are also utilized with a fine-tuning strategy in this work. Besides, each model has a different depth. As the depth of the model increases, more abstract features containing the distinguishing information are extracted by the model. In this work, the best classification performance among the single models is achieved with the deepest lightweight model, i.e. fine-tuned EfficientNet-B0. However, the model depth may cause some semantic features to dampen. Therefore, in order to compensate for the loss of semantic information in the deepest model, the fine-tuned SqueezeNet with 22-layer of depth, and the fine-tuned ShuffleNet with 50-layer of depth are combined with the fine-tuned EfficienNet-B0 with 82-layer of depth. Experimental results confirm that combining these three models compensates for the loss of information. Moreover, the best results are achieved with the lightweight ensemble network model. We believe that the proposed method in this work may be useful for clinicians or medical experts.

### References

[1] W. Yang *et al.*, 'The role of imaging in 2019 novel coronavirus pneumonia (COVID-19)', *Eur. Radiol.*, vol. 30, no. 9, pp. 4874–4882, Sep. 2020, doi: 10.1007/s00330-020-06827-4.

[2] 'WHO Director-General's opening remarks at the media briefing on COVID-19 - 11 March 2020'. https://www.who.int/director-general/speeches/detail/who-director-general-s-opening-remarks-at-the-media-briefing-on-covid-19---11-march-2020 (accessed Nov. 16, 2020).

[3] 'WHO Coronavirus Disease (COVID-19) Dashboard'. https://covid19.who.int (accessed Dec. 05, 2020).

[4] 'Draft landscape of COVID-19 candidate vaccines'. https://www.who.int/publications/m/item/draft-landscape-of-covid-19-candidate-vaccines (accessed Nov. 27, 2020).

[5] H. Ledford, D. Cyranoski, and R. V. Noorden, 'The UK has approved a COVID vaccine — here's what scientists now want to know', *Nature*, Dec. 2020, doi: 10.1038/d41586-020-03441-8.

[6] R. Frutos and C. A. Devaux, 'Mass culling of minks to protect the COVID-19 vaccines: is it rational?', *New Microbes New Infect.*, vol. 38, p. 100816, Nov. 2020, doi: 10.1016/j.nmni.2020.100816.

[7] 'Transmission of SARS-CoV-2: implications for infection prevention precautions'. https://www.who.int/news-room/commentaries/detail/transmission-of-sars-cov-2-implications-for-infection-prevention-precautions (accessed Nov. 24, 2020).

[8] C. Li, C. Zhao, J. Bao, B. Tang, Y. Wang, and B. Gu, 'Laboratory diagnosis of coronavirus disease-2019 (COVID-19)', *Clin. Chim. Acta*, vol. 510, pp. 35–46, Nov. 2020, doi: 10.1016/j.cca.2020.06.045.

[9] 'Coronavirus'. https://www.who.int/westernpacific/health-topics/coronavirus (accessed Nov. 26, 2020).

[10] T. Struyf *et al.*, 'Signs and symptoms to determine if a patient presenting in primary care or hospital outpatient settings has COVID-19 disease', *Cochrane Database Syst. Rev.*, no. 7, 2020, doi: 10.1002/14651858.CD013665.

[11] T. Ji *et al.*, 'Detection of COVID-19: A review of the current literature and future perspectives', *Biosens. Bioelectron.*, vol. 166, p. 112455, Oct. 2020, doi: 10.1016/j.bios.2020.112455.

[12] M. E. H. Chowdhury *et al.*, 'Can AI Help in Screening Viral and COVID-19 Pneumonia?', *IEEE Access*, vol. 8, pp. 132665–132676, 2020, doi: 10.1109/ACCESS.2020.3010287.

[13] T. C. Williams *et al.*, 'Sensitivity of RT-PCR testing of upper respiratory tract samples for SARS-CoV-2 in hospitalised patients: a retrospective cohort study', *Wellcome Open Res.*, vol. 5, p. 254, Oct. 2020, doi: 10.12688/wellcomeopenres.16342.1.

[14] Z. Y. Zu *et al.*, 'Coronavirus Disease 2019 (COVID-19): A Perspective from China', *Radiology*, vol. 296, no. 2, pp. E15–E25, Aug. 2020, doi: 10.1148/radiol.2020200490.

[15] T. C. H. Hui *et al.*, 'Clinical utility of chest radiography for severe COVID-19', *Quant. Imaging Med. Surg.*, vol. 10, no. 7, pp. 1540–1550, Jul. 2020, doi: 10.21037/qims-20-642.

[16] K. Ahammed, Md. S. Satu, M. Z. Abedin, Md. A. Rahaman, and S. M. S. Islam, 'Early Detection of Coronavirus Cases Using Chest X-ray Images Employing Machine Learning and Deep Learning Approaches', Infectious Diseases (except HIV/AIDS), preprint, Jun. 2020. doi: 10.1101/2020.06.07.20124594.

[17] T. Ozturk, M. Talo, E. A. Yildirim, U. B. Baloglu, O. Yildirim, and U. Rajendra Acharya, 'Automated detection of COVID-19 cases using deep neural networks with X-ray images', *Comput. Biol. Med.*, vol. 121, p. 103792, Jun. 2020, doi: 10.1016/j.compbiomed.2020.103792.

[18] A. M. Ismael and A. Şengür, 'Deep learning approaches for COVID-19 detection based on chest X-ray images', *Expert Syst. Appl.*, vol. 164, p. 114054, Feb. 2021, doi: 10.1016/j.eswa.2020.114054.

[19] O. Russakovsky *et al.*, 'ImageNet Large Scale Visual Recognition Challenge', *ArXiv14090575 Cs*, Jan. 2015, Accessed: Nov. 17, 2020. [Online]. Available: http://arxiv.org/abs/1409.0575.

[20] S. Minaee, R. Kafieh, M. Sonka, S. Yazdani, and G. Jamalipour Soufi, 'Deep-COVID: Predicting COVID-19 from chest X-ray images using deep transfer learning', *Med. Image Anal.*, vol. 65, p. 101794, Oct. 2020, doi: 10.1016/j.media.2020.101794.

[21] M. Nour, Z. Cömert, and K. Polat, 'A Novel Medical Diagnosis model for COVID-19 infection detection based on Deep Features and Bayesian Optimization', *Appl. Soft Comput.*, p. 106580, Jul. 2020, doi: 10.1016/j.asoc.2020.106580.

[22] R. Karthik, R. Menaka, and H. M., 'Learning distinctive filters for COVID-19 detection from chest X-ray using shuffled residual CNN', *Appl. Soft Comput.*, p. 106744, Sep. 2020, doi: 10.1016/j.asoc.2020.106744.

[23] M. Farooq and A. Hafeez, 'COVID-ResNet: A Deep Learning Framework for Screening of COVID19 from Radiographs', *ArXiv200314395 Cs Eess*, Mar. 2020, Accessed: Nov. 25, 2020. [Online]. Available: http://arxiv.org/abs/2003.14395.

[24] L. Wang and A. Wong, 'COVID-Net: A Tailored Deep Convolutional Neural Network Design for Detection of COVID-19 Cases from Chest X-Ray Images', *ArXiv200309871 Cs Eess*, May 2020, Accessed: Nov. 25, 2020. [Online]. Available: http://arxiv.org/abs/2003.09871.

[25] A. Mangal *et al.*, 'CovidAID: COVID-19 Detection Using Chest X-Ray', *ArXiv200409803 Cs Eess*, Apr. 2020, Accessed: Nov. 25, 2020. [Online]. Available: http://arxiv.org/abs/2004.09803.

[26] A. Abbas, M. M. Abdelsamea, and M. M. Gaber, 'Classification of COVID-19 in chest X-ray images using DeTraC deep convolutional neural network', *ArXiv200313815 Cs Eess Stat*, May 2020, Accessed: Nov. 25, 2020. [Online]. Available: http://arxiv.org/abs/2003.13815.

[27] L. Brunese, F. Mercaldo, A. Reginelli, and A. Santone, 'Explainable Deep Learning for Pulmonary Disease and Coronavirus COVID-19 Detection from X-rays', *Comput. Methods Programs Biomed.*, vol. 196, p. 105608, Nov. 2020, doi: 10.1016/j.cmpb.2020.105608.

[28] C. Ouchicha, O. Ammor, and M. Meknassi, 'CVDNet: A novel deep learning architecture for detection of coronavirus (Covid-19) from chest x-ray images', *Chaos Solitons Fractals*, vol. 140, p. 110245, Nov. 2020, doi: 10.1016/j.chaos.2020.110245.

[29] F. A. Mettler, W. Huda, T. T. Yoshizumi, and M. Mahesh, 'Effective Doses in Radiology and Diagnostic Nuclear Medicine: A Catalog', *Radiology*, vol. 248, no. 1, pp. 254–263, Jul. 2008, doi: 10.1148/radiol.2481071451.

[30] F. N. Iandola, S. Han, M. W. Moskewicz, K. Ashraf, W. J. Dally, and K. Keutzer, 'SqueezeNet: AlexNet-level accuracy with 50x fewer parameters and <0.5MB model size', *ArXiv160207360 Cs*, Nov. 2016, Accessed: Nov. 17, 2020. [Online]. Available: http://arxiv.org/abs/1602.07360.

[31] X. Zhang, X. Zhou, M. Lin, and J. Sun, 'ShuffleNet: An Extremely Efficient Convolutional Neural Network for Mobile Devices', *ArXiv170701083 Cs*, Dec. 2017, Accessed: Nov. 17, 2020. [Online]. Available: http://arxiv.org/abs/1707.01083.

[32] M. Sandler, A. Howard, M. Zhu, A. Zhmoginov, and L.-C. Chen, 'MobileNetV2: Inverted Residuals and Linear Bottlenecks', *ArXiv180104381 Cs*, Mar. 2019, Accessed: Nov. 18, 2020. [Online]. Available: http://arxiv.org/abs/1801.04381.

[33] M. Tan and Q. V. Le, 'EfficientNet: Rethinking Model Scaling for Convolutional Neural Networks', *ArXiv190511946 Cs Stat*, Sep. 2020, Accessed: Nov. 17, 2020. [Online]. Available: http://arxiv.org/abs/1905.11946.

[34] C. Tan, F. Sun, T. Kong, W. Zhang, C. Yang, and C. Liu, 'A Survey on Deep Transfer Learning', in *Artificial Neural Networks and Machine Learning – ICANN 2018*, Cham, 2018, pp. 270–279, doi: 10.1007/978-3-030-01424-7_27.

[35] C. Öksüz and M. K. Güllü, 'Deep Feature Extraction Based Fine-Tuning', presented at the The 28th IEEE Conference on Signal Processing and Communications Applications, Gaziantep, Turkey, Oct. 2020.

[36] A. Krizhevsky, I. Sutskever, and G. E. Hinton, 'ImageNet classification with deep convolutional neural networks', *Commun. ACM*, vol. 60, no. 6, pp. 84–90, May 2017, doi: 10.1145/3065386.

[37] R. R. Selvaraju, M. Cogswell, A. Das, R. Vedantam, D. Parikh, and D. Batra, 'Grad-CAM: Visual Explanations from Deep Networks via Gradient-Based Localization', in *2017 IEEE International Conference on Computer Vision (ICCV)*, Venice, Oct. 2017, pp. 618–626, doi: 10.1109/ICCV.2017.74.

[38] J. P. Cohen, *ieee8023/covid-chestxray-dataset*. 2020.

[39] D. S. Kermany *et al.*, 'Identifying Medical Diagnoses and Treatable Diseases by Image-Based Deep Learning', *Cell*, vol. 172, no. 5, pp. 1122-1131.e9, Feb. 2018, doi: 10.1016/j.cell.2018.02.010.

[40] B. G. S. Cruz, J. Sölter, M. N. Bossa, and A. D. Husch, 'On the Composition and Limitations of Publicly Available COVID-19 X-Ray Imaging Datasets', *ArXiv200811572 Cs Eess*, Aug. 2020, Accessed: Dec. 07, 2020. [Online]. Available: http://arxiv.org/abs/2008.11572.

[41] S. R. Nayak, D. R. Nayak, U. Sinha, V. Arora, and R. B. Pachori, 'Application of deep learning techniques for detection of COVID-19 cases using chest X-ray images: A comprehensive study', *Biomed. Signal Process. Control*, vol. 64, p. 102365, Feb. 2021, doi: 10.1016/j.bspc.2020.102365.

[42] A. Altan and S. Karasu, 'Recognition of COVID-19 disease from X-ray images by hybrid model consisting of 2D curvelet transform, chaotic salp swarm algorithm and deep learning technique', *Chaos Solitons Fractals*, vol. 140, p. 110071, Nov. 2020, doi: 10.1016/j.chaos.2020.110071.